\documentclass[journal]{IEEEtran}
\IEEEoverridecommandlockouts
\usepackage{cite}
\usepackage{amsmath,amssymb,amsfonts}
\usepackage{algorithmic}
\usepackage{graphicx}
\usepackage{textcomp}
\usepackage{xcolor}
\usepackage{amsthm}
\usepackage{enumitem}
\usepackage{subfig}
\usepackage[ruled,linesnumbered]{algorithm2e}
\usepackage{balance}
\usepackage{changes}
\usepackage{cancel}

\allowdisplaybreaks
\theoremstyle{plain}
\newtheorem{assumption}{Assumption}
\theoremstyle{remark}
\newtheorem{remark}{Remark}

\begin{document}

\title{Distributed Online Feedback Optimization for Real-time Distribution System Voltage Regulation}

\author{Sen~Zhan,~\IEEEmembership{Member,~IEEE,}
        Nikolaos~G.~Paterakis,~\IEEEmembership{Senior Member,~IEEE,}\\
        Wouter~van~den~Akker,~\IEEEmembership{Member,~IEEE,}
        Anne~van~der~Molen,~\IEEEmembership{Member,~IEEE,}\\
        Johan~Morren,~\IEEEmembership{Member,~IEEE,}
        and~Han~Slootweg,~\IEEEmembership{Senior Member,~IEEE}
\thanks{The authors are with the Department of Electrical Engineering, Eindhoven University of Technology, 5600 MB Eindhoven, The Netherlands (e-mail: s.zhan@tue.nl; n.paterakis@tue.nl;  w.f.v.d.akker@tue.nl; a.e.v.d.molen@tue.nl; j.morren@tue.nl; j.g.slootweg@tue.nl).}%
\thanks{W. van den Akker is also with the Corporate Strategy Department, Alliander, 6812 AH Arnhem, The Netherlands.}%
\thanks{A. van der Molen is also with the Grid Strategy Department, Stedin, 3011 TA Rotterdam, The Netherlands.}%
\thanks{J. Morren and H. Slootweg are also with the Department of Asset Management, Enexis, 5223 MB 's-Hertogenbosch, The Netherlands.}%
\thanks{This work was supported by the TKI Urban Energy from the `Toeslag voor Topconsortia voor Kennis en Innovatie (TKI)' of the Ministry of Economic Affairs and Climate Policy under Grant 1821401. This work is also part of the NO-GIZMOS project (MOOI52109) which received funding from the Topsector Energie MOOI subsidy program of the Netherlands Ministry of Economic Affairs and Climate Policy, executed by the Netherlands Enterprise Agency (RVO).}
}

\onecolumn
This article has been accepted for publication in IEEE Transactions on Power Systems. This is the author's version which has not been fully edited and
content may change prior to final publication. Citation information: DOI 10.1109/TPWRS.2025.3585607

\textcopyright  2025 IEEE. All rights reserved, including rights for text and data mining and training of artificial intelligence and similar technologies. Personal use is permitted, but
republication/redistribution requires IEEE permission. See https://www.ieee.org/publications/rights/index.html for more information.

Link to IEEEXplore: https://ieeexplore.ieee.org/document/11066276

\twocolumn
\maketitle

\begin{abstract}
We investigate the real-time voltage regulation problem in distribution systems employing online feedback optimization (OFO) with short-range communication between physical neighbours. OFO does not need an accurate grid model nor estimated consumption of non-controllable loads, affords fast calculations, and demonstrates robustness to uncertainties and disturbances, which render it particularly suitable for real-time distribution system applications. However, many OFO controllers require centralized communication, making them susceptible to single-point failures. This paper proposes a distributed OFO design based on a nested feedback optimization strategy and analyzes its convergence. The strategy preserves end-users' privacy by keeping voltage data local. Numerical study results demonstrate that the proposed design achieves effective voltage regulation and outperforms other distributed and local approaches.\looseness=-1

\end{abstract}

\begin{IEEEkeywords}
Distributed communication, online feedback optimization, reactive power control, voltage regulation  
\end{IEEEkeywords}

\section{Introduction}

The rapid growth of distributed energy resources (DERs) has introduced significant uncertainty and volatility to distribution systems. With the high resistance/reactance ratios of distribution cables, voltage fluctuations are occurring more frequently. Consequently, real-time voltage regulation has become a significant operational challenge for distribution system operators (DSOs). The conventional approach of DSOs is grid reinforcement, but this demands significant investments, a skilled workforce, and considerable time. During this lengthy process, a complementary solution is to exploit the flexibility of network assets and DERs. In this context, online feedback optimization (OFO) has recently emerged as a promising strategy, e.g. in \cite{DallAnese2018,Bernstein2019,Bolognani2019,Ortmann2024,Guo2023,Bolognani2015}.

\begin{figure}[t]
  \centering
  \includegraphics[width=.85\columnwidth]{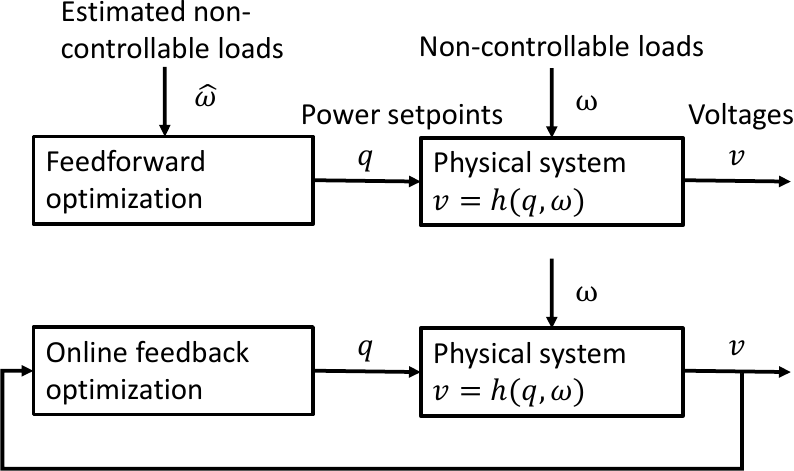}
  \caption{Block diagrams of feedforward and feedback systems for voltage regulation using reactive power, where $q$ represents system input (reactive power), $v$ represents system output (voltages), $\omega$ ($\widehat{\omega}$) represents (estimated) system disturbance (non-controllable loads), and $h$ maps system input and disturbance to system output. This figure is adapted from \cite{Ortmann2023}.}
  \label{fig:feedback}
\end{figure}

OFO utilizes measurements as feedback and employs optimization algorithms as feedback controllers to steer physical systems towards their optimal operating points \cite{Hauswirth2021}, which are defined by the well-established optimal power flow (OPF) problem. Compared to directly solving the OPF problem in a feedforward manner, OFO does not need an accurate grid model nor data of non-controllable loads, affords fast calculations, and is robust against uncertainties and disturbances in distribution systems due to its feedback-based nature \cite{Haberle2021,Hauswirth2021,Ortmann2020}.  Figure \ref{fig:feedback} illustrates feedforward and feedback systems.

OFO also brings distinct advantages over other existing approaches. Compared to local droop control \cite{Li2014,Bolognani2019,Ortmann2020_dist}, OFO ensures more reliable voltage regulation \cite{Bolognani2019} and can pursue grid-level objectives. In contrast to deep reinforcement learning \cite{Cao2021,Zhang2021}, OFO does not need a complicated (often centralized offline) training process while offering theoretical guarantees to obtain globally optimal solutions. Recent studies in \cite{Gupta2024,Yuan2024,Yuan20242,Cui2022} have explored data-driven approaches to train linear and nonlinear local voltage controllers via deep learning, reinforcement learning, and unsupervised learning. However, these approaches demand extensive training data and lack optimality guarantees without real-time coordination between DERs. Furthermore, unlike using traditional voltage regulators such as on-load tap changers and capacitor banks \cite{Nazir2021}, OFO provides rapid voltage regulation with inverter-interfaced DERs and can complement these slow-reacting devices in real \mbox{time \cite{Tang2021}}. \looseness=-1


While many OFO controllers rely on centralized communication \cite{DallAnese2018,DallAnese2018_vpp, Bernstein2019,Haberle2021,Ortmann2024,Guo2023,Zhan2023,Ortmann2020,Zhou2018,Zhan2024,Cave2024}, distributed OFO \mbox{\cite{Bolognani2015,Bolognani2019,Tang2021,Magnusson2020,Tang2024,Qu2020}} based on communication between physical neighbours appears to be a compelling approach due to its robustness to single-point failures and enhancement of privacy. Among the existing distributed designs, \cite{Bolognani2015,Bolognani2019,Tang2021} demand particular forms of objective functions, while \cite{Magnusson2020} relies on disseminating global information through neighbouring nodes. However, the systematic delays inherent in this information propagation may raise stability issues, potentially undermining its efficacy in practical distribution systems. In \cite{Tang2024}, the distributed design was achieved by only keeping components in the network sensitivity matrix related to local and neighbouring nodes. This heuristic approach does not assure adequate voltage regulation.

The motivation of this study is to develop a real-time voltage regulation strategy that optimally coordinates DERs on a fast timescale and operates without requiring an accurate grid model or load data. Furthermore, the strategy should be robust to single-point failures, meaning that unlike centralized approaches—where the failure of a single coordination unit can disrupt the entire system—the approach should ensure that voltage regulation remains functional even if some nodes fail. Finally, the strategy should preserve privacy. Exposing nodal voltage data could allow external entities to infer sensitive load information, which is typically considered private.

In this paper, we propose a distributed OFO design that meets these requirements. The strategy is inspired by \cite{Qu2020}, where the distributed implementation was enabled by the sparsity of the inverse of the network sensitivity matrix $\mathbf{X}$, i.e. $\mathbf{X}^{-1}$ is sparse. Both our proposed and their approaches continue with scaling the gradient by $\mathbf{X}^{-1}$ followed by a projection step to ensure satisfaction of the DER capacity constraint. It is conceptually attractive to use the Euclidean distance in the projection, which can be implemented using only local information. However, as shown in  \cite{Bertsekas1999}, the distance measured by the norm $\lVert \mathbf{z} \rVert _\mathbf{X} = \sqrt{\mathbf{z}^\intercal \mathbf{X} \mathbf{z}} $ should be used instead in the projection step to ensure descent iterations and thus convergence. With the Euclidean distance, it would be a \textit{two-metric} approach which can lead to algorithm \mbox{divergence \cite{Bertsekas1999}}. Solving this non-Euclidean projection problem would however require global information since $\mathbf{X}$ is dense, contradicting the distributed OFO design. As a remedy, \cite{Qu2020} introduced additional dualization of the DER capacity constraint on top of the Euclidean projection to ensure the DER constraint satisfaction. Nevertheless, this strategy takes a prohibitively large number of iterations to converge, which may jeopardize its performance in online implementation.





We tackle this challenge by proposing an iterative projected gradient descent algorithm to solve this non-Euclidean projection problem where each node only requires information from its neighbours. This leads to our overall nested feedback optimization approach, where the outer loop comprises the OFO iterations requiring communication between physical neighbours and yielding tentative but not necessarily feasible DER setpoints, while the inner loop solves the non-Euclidean projection problem to map the tentative setpoints to actual feasible setpoints. Unlike the centralized approaches, our method is robust to single-point failures as it operates without a central coordination unit and preserves privacy. Compared to \cite{Bolognani2019,Bolognani2015,Tang2021}, our approach accommodates a broader range of objective functions. In contrast to \cite{Magnusson2020,Tang2024}, our approach keeps voltage information local enhancing privacy. Finally, compared to \cite{Qu2020}, our approach converges in significantly fewer iterations and achieves superior voltage regulation performance. To summarize, the main contributions of this paper are:
\begin{itemize}
  \item We develop a projected gradient descent algorithm that exchanges information between neighboring nodes in a backward-forward manner to solve the non-Euclidean projection problem.
  \item We propose a nested feedback optimization strategy for distribution system voltage regulation that operates without requiring an accurate grid model or load data. The strategy optimally coordinates DERs on a fast timescale while ensuring robustness to single-point failures and preserving privacy by keeping voltage data local.
  \item We theoretically prove the convergence of the proposed strategy. Simulation results demonstrate that it achieves comparable voltage regulation to its centralized counterpart while significantly outperforming other distributed and local approaches.
\end{itemize}

The remainder of this paper is structured as follows: \mbox{Section \ref{sec:method} } presents the system modeling, problem formulation, and a centralized OFO strategy. Section \ref{sec:dist} introduces the proposed nested approach and analyzes its convergence. Section \ref{sec:sim} presents a numerical study, while Section \ref{sec:concl} draws conclusions and discusses future work.

\section{Problem formulation and centralized OFO}\label{sec:method}
\subsection{Distribution system modeling}
Consider a balanced radial distribution system with \mbox{$N+1$} buses collected in the set $\mathcal{N}=\{0,1,\cdots,N\}$, and cables collected in the set $\mathcal{E}=\{(i,j)\}\in\mathcal{N}\times\mathcal{N}$. Bus $0$ is the secondary substation bus and is assumed to have a fixed voltage magnitude $v_0$. A cable is denoted by $(i,j)$ if bus $i$ is closer to the substation bus than bus $j$. Define $\mathcal{N}^+=\mathcal{N} \symbol{92} \{0\}$. For each bus $i \in \mathcal{N}^+$, let $p_i$ and $q_i$ be the active and reactive power injections from the DER, let $p_i^d$ and $q_i^d$ be the active and reactive power demand, respectively, and let $v_i$ be the voltage magnitude. For each cable $(i,j)\in\mathcal{E}$, let $P_{ij}$ and $Q_{ij}$ be the active and reactive power flows from buses $i$ to $j$, and let $r_{ij}$ and $x_{ij}$ be its resistance and reactance, respectively. Finally, let bold uppercase and lowercase letters denote matrices and column vectors, respectively, with components defined earlier, e.g. $\mathbf{v} = [v_1,v_2,\cdots,v_N]^\intercal$.

The linearized \textit{DistFlow} equations in (\ref{lpf}) were proposed in \cite{Baran1989} and can be used to model balanced radial distribution systems. Equations (\ref{lpf:p})-(\ref{lpf:q}) represent nodal active and reactive power balance constraints, respectively. The voltage relation is modeled in (\ref{lpf:v}), which further leverages the assumption that $v_i^2-v_j^2 \approx 2 (v_i-v_j)$ since $v_i \approx 1 \text{ pu}, \forall i \in \mathcal{N}$.
\begin{subequations}\label{lpf}
\begin{align}
  \label{lpf:p} & P_{ij} + p_j = \sum_{k:(j,k)\in\mathcal{E}} P_{jk} + p^d_j, \forall j \in \mathcal{N}^+,\\
  \label{lpf:q} & Q_{ij} + q_j = \sum_{k:(j,k)\in\mathcal{E}} Q_{jk} + q^d_j, \forall j \in \mathcal{N}^+,\\
  \label{lpf:v} & v_i - v_j = r_{ij} P_{ij} + x_{ij} Q_{ij}, \forall (i,j) \in \mathcal{E}.
\end{align}
\end{subequations}

Applying (\ref{lpf}) directly in a feedforward voltage regulation scheme requires estimates of $p_j^d$ and $q^d_j$, $\forall j \in \mathcal{N}^+$. Moreover, the modeling inaccuracy will also result in voltage approximation errors. To synthesize the proposed feedback controller for voltage regulation, we leverage (\ref{lpf}) to derive network sensitivities, which relate changes in nodal active and reactive power injections to voltage variations. Following the steps in \cite{Farivar2013} or \cite{Zhu2016}, the linear relation in (\ref{lpf:vcomp}) can be constructed.  
\begin{align} \label{lpf:vcomp}
  \mathbf{v} = v_0 \mathbf{1} + \mathbf{R} (\mathbf{p}-\mathbf{p}^d) + \mathbf{X} (\mathbf{q}-\mathbf{q}^d).
\end{align}

The $N\times N$-dimensional symmetrical matrices $\mathbf{R}$ and $\mathbf{X}$ are given in (\ref{sensitivities_components}), where $\mathcal{E}_i$ represents the set of cables on the unique path from the substation bus to bus $i$.
\begin{align} \label{sensitivities_components}
  R_{ij} = \sum_{(h,k)\in \mathcal{E}_i \cap \mathcal{E}_j} r_{hk},   X_{ij} = \sum_{(h,k)\in \mathcal{E}_i \cap \mathcal{E}_j} x_{hk}.
\end{align}

These matrices accordingly capture the network sensitivities, i.e.
\begin{align} \label{sensitivities}
  \frac{\partial \mathbf{v}}{\partial \mathbf{p}} = \mathbf{R}, \frac{\partial \mathbf{v}}{\partial \mathbf{q}} = \mathbf{X}.
\end{align}

\begin{figure}[t]
  \centering
  \subfloat[$\mathbf{X}$ (pu/kVar)]{\includegraphics[width=0.9\linewidth]{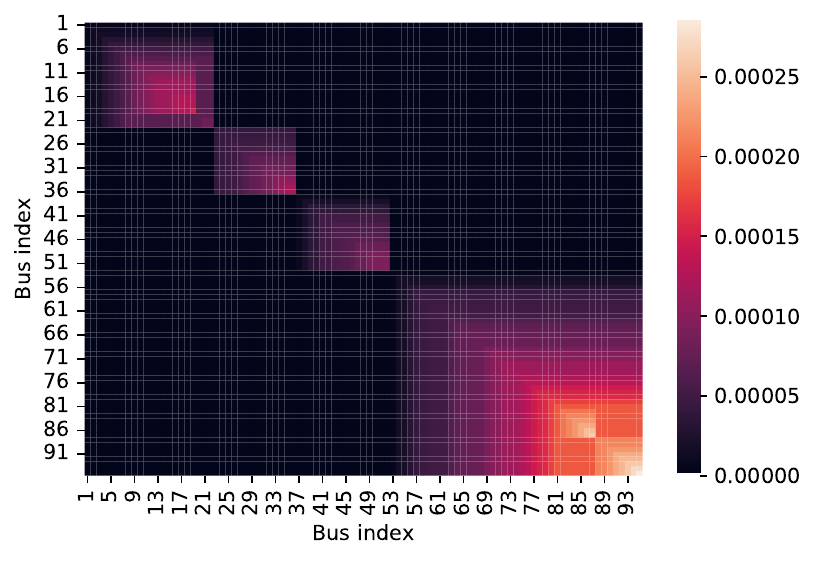}}\\\vspace{-10pt}
  \subfloat[$\mathbf{X}^{-1}$ (kVar/pu)]{\includegraphics[width=0.9\linewidth]{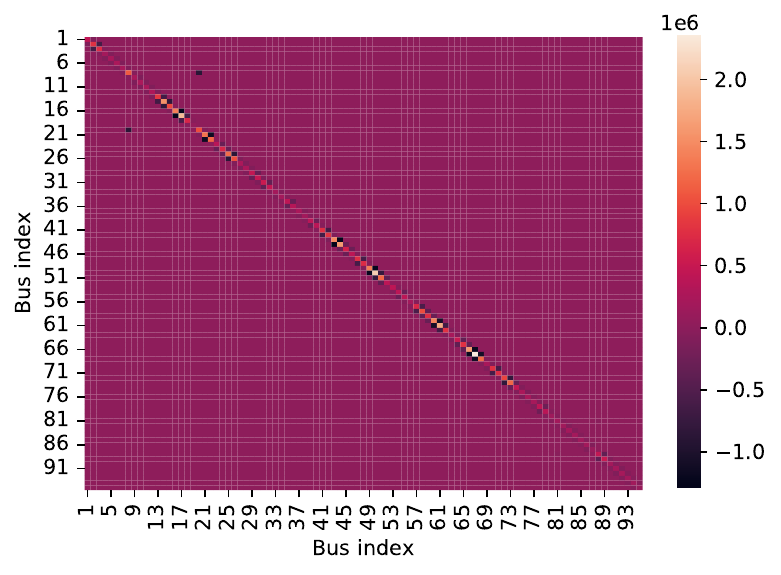}}
  \caption{Heat maps of the network sensitivity matrix $\mathbf{X}$ and its inverse $\mathbf{X}^{-1}$.}
  \label{fig:heatmap}
\end{figure}

When the resistances and reactances of the cables are all positive, \cite{Farivar2013} shows that $\mathbf{R}$ and $\mathbf{X}$ are both symmetrical positive definite. Furthermore, \cite{Qu2020} shows that their inverse matrices $\mathbf{R}^{-1}$ and $\mathbf{X}^{-1}$ have the sparsity pattern such that
\begin{align} \label{sparsity}
  R_{ij}^{-1} \neq 0, X_{ij}^{-1} \neq 0 \iff (i,j) \in \mathcal{E} \text{ or } i = j.
\end{align}
Figure \ref{fig:heatmap} visualizes $\mathbf{X}$ and $\mathbf{X}^{-1}$ for the test system in \mbox{Section \ref{sec:sim}}, where the sparsity patterns are shown.

\subsection{Problem formulation}
Consider the problem of distribution system voltage regulation using reactive power as formulated in (\ref{opt}). The objective function (\ref{obj}) minimizes the reactive power use cost, where $c_i > 0$ is the cost parameter. Constraint (\ref{cons:voltage}) enforces the lower and upper voltage limits $\underline{v}$ and $\overline{v}$, where $\mu_i$ and $\lambda_i$ are the respective dual variables. Finally, (\ref{cons:reactive}) ensures that the reactive power injection $q_i$ respects the lower and upper limits $\underline{q}_i$ and $\overline{q}_i$, which capture DER capacity and power factor constraints. It is noteworthy that $v_i$ is dependent on the decision variable $q_i$ and other disturbances in the system such as active and reactive power consumption of non-controllable loads. 
\begin{subequations}\label{opt}
\begin{align}
   \underset{q_i,\forall i \in \mathcal{N}^+}{\operatorname{minimize }} & \quad \sum_{i\in\mathcal{N}^+} \frac{1}{2} c_i q_i^2, \label{obj}\\
   \operatorname{s.t. } & \quad \underline{v} \leq v_i \leq \overline{v}: \mu_i, \lambda_i, \forall i \in \mathcal{N}^+ ,\label{cons:voltage}\\
   & \underline{q}_i \leq q_i \leq \overline{q}_i,  \forall i \in \mathcal{N}^+. \label{cons:reactive}
\end{align}   
\end{subequations}

  \begin{remark}
    The objective function is designed to enable our distributed voltage regulation algorithm. Since the algorithm relies solely on local information and data from neighboring nodes, the gradient of the objective function must be locally computable. This requirement excludes certain objectives, such as loss minimization, which would require a centralized communication structure. Additionally, in case of overvoltage due to PV generation, minimizing reactive power usage contributes to loss reduction. When PV inverters absorb reactive power to lower voltage magnitudes, they increase reactive power transfer within the system, which already experiences reactive power consumption from non-controllable loads and cable losses. The more reactive power PV inverters absorb, the higher the system losses. Thus, minimizing their reactive power consumption helps reduce overall losses, as demonstrated by numerical results in Section \ref{sec:result-static}.
    Finally, this objective function also enhances inverter reliability. As noted in \cite{pvlifetime}, operating inverters at power factors significantly different from unity can reduce their lifespan. Therefore, our approach aligns with the interest of end-users by promoting longer inverter lifetime.
  \end{remark}

For further developments, the dual problem of (\ref{opt}) is formulated in (\ref{dual}).
\begin{align}
  \label{dual}& \underset{\mu_i \geq 0, \lambda_i\geq 0, \forall i \in \mathcal{N}^+}{\operatorname{maximize }} \big\{  \underset{\underline{q}_i \leq q_i \leq \overline{q}_i,,\forall i \in \mathcal{N}^+}{\operatorname{minimize }} \quad \mathcal{L}(\mathbf{q}, \boldsymbol{\mu}, \boldsymbol{\lambda}) \big \},
\end{align}
where the partial Lagrangian function is defined in (\ref{lagrangian}).
\begin{align}\label{lagrangian}
  \mathcal{L}(\mathbf{q}, \boldsymbol{\mu}, \boldsymbol{\lambda})=   \sum_{i \in \mathcal{N}^{+}} \left[ \frac{1}{2} c_i q_i^2 + \lambda_i\left(v_i-\overline{v}\right) + \mu_i\left(\underline{v}-v_i\right) \right].
\end{align}

\subsection{Centralized primal-dual gradient projection algorithm}
Solving (\ref{opt}) offline in a feedforward way demands an accurate grid model and disturbance data to evaluate $v_i$, which might not be readily available. The feedforward approach also lacks robustness \cite{Hauswirth2021}. Thus, this section introduces a feedback controller based on OFO. The controller leverages primal-dual gradient projection (PDGP) as the underlying algorithm \cite{DallAnese2018,DallAnese2018_vpp,Bernstein2019} to track the time-varying optimizer of (\ref{opt}), which requires a centralized gather-and-broadcast communication architecture. Compared to other dual optimization approaches such as dual ascent and the alternating method of multipliers (ADMM), PDGP offers additional tuning of primal step sizes, which is beneficial under fluctuating grid conditions, and enables faster implementation of DER setpoints. The centralized OFO-based solution is included primarily as a baseline for comparison and as a foundation for the distributed methods developed later.

Specifically, the PDGP algorithm performs projected gradient ascent and descent steps iteratively for the dual and primal variables, respectively. At each time step (iteration) $k$, it repeats the following three steps:


\begin{enumerate}[leftmargin=*]
  \item For each bus $i \in \mathcal{N}^+$, collect its voltage measurement $\tilde{v}_i^k$ and update $\lambda_i$ and $\mu_i$ using (\ref{grad:lambda})-(\ref{grad:mu}), where the projection operator is defined as $[u]^+= \max(u,0)$, $\alpha$ with various superscripts represents step sizes, while $r^d$ and $r^p$ in (\ref{grad:centralized}) are the dual and primal regularization factors \cite{DallAnese2018,DallAnese2018_vpp,Bernstein2019}, respectively.
  \begin{subequations}\label{pdgp}
  \begin{align}
     & \lambda_i^{k+1} = \big[\lambda_i^{k} + \alpha^{d} (\tilde{v}_i^k - \overline{v}-r^d \lambda_i^{k}) \big]^+ ,\label{grad:lambda}\\
     & \mu_i^{k+1} =  \big[\mu_i^{k} + \alpha^{d} (\underline{v}-\tilde{v}_i^k -r^d \mu_i^{k}) \big]^+. \label{grad:mu}
  \end{align}
\item The DSO broadcasts the updated dual variables $\boldsymbol{\lambda}^{k+1}$ and $\boldsymbol{\mu}^{k+1}$ to all buses.
\item For each bus $i\in\mathcal{N}^+$, update its reactive power setpoints $q_i$ using (\ref{grad:centralized}), where $[u]_{\underline{u}}^{\overline{u}}$ represents the Euclidean projection into $[\underline{u}, \overline{u}]$. Since $\mathbf{X}$ is dense, \mbox{Step 2} is needed to provide information for Step 3. 
\begin{equation}\label{grad:centralized}
 q_i^{k+1} = \big[q_i^k - \alpha [c_i q_i^k + [\mathbf{X} (\boldsymbol{\lambda}^{k+1}-\boldsymbol{\mu}^{k+1}+ r^p \mathbf{q}^k)]_i ]  \big]_{\underline{q}_i^k}^{\overline{q}_i^k}.
\end{equation}
\end{subequations}
\end{enumerate}

  \begin{remark}\label{re:reg}
The regularization ensures strong convexity of the problem and guarantees convergence. However, it introduces a deviation from the exact solution of (\ref{dual}), resulting in a close approximation instead. As shown in \cite{Koshal2011}, the discrepancy is bounded and proportional to $\sqrt{r_d}$. A practical guide for choosing the regularization parameters is to start with small values and gradually increase them until the algorithm stabilizes.
  \end{remark}

\section{Distributed OFO} \label{sec:dist}
\subsection{Two-metric distributed approach}\label{sec:two-metric}
To avoid the centralized communication requirement and enable the distributed communication architecture where communication is only between physical neighbours, \cite{Qu2020} proposed to scale the gradient in (\ref{grad:centralized}) by the sparse positive definite matrix $\mathbf{X}^{-1}$, which has non-zero entries only in the diagonal and the $i$-th row $j$-th column if buses $i$ and $j$ are adjacent, i.e. as in (\ref{sparsity}). Continuing with the Euclidean projection yields (\ref{grad:two-metric}) which replaces (\ref{grad:centralized}) in the OFO, where $\mathbf{C}$ is a $N$-dimensional diagonal matrix with $c_i$ defined in (\ref{obj}). 
\begin{equation}\label{grad:two-metric}
    q_i^{k+1} = \big[q_i^k - \alpha [[\mathbf{X}^{-1} \mathbf{C} \mathbf{q}^k]_i + \lambda_i^{k+1}-\mu_i^{k+1} + r^p q_i^k] \big]_{\underline{q}_i^k}^{\overline{q}_i^k}.
   \end{equation}

\begin{figure}[t]
  \centering
  \includegraphics[width=\linewidth]{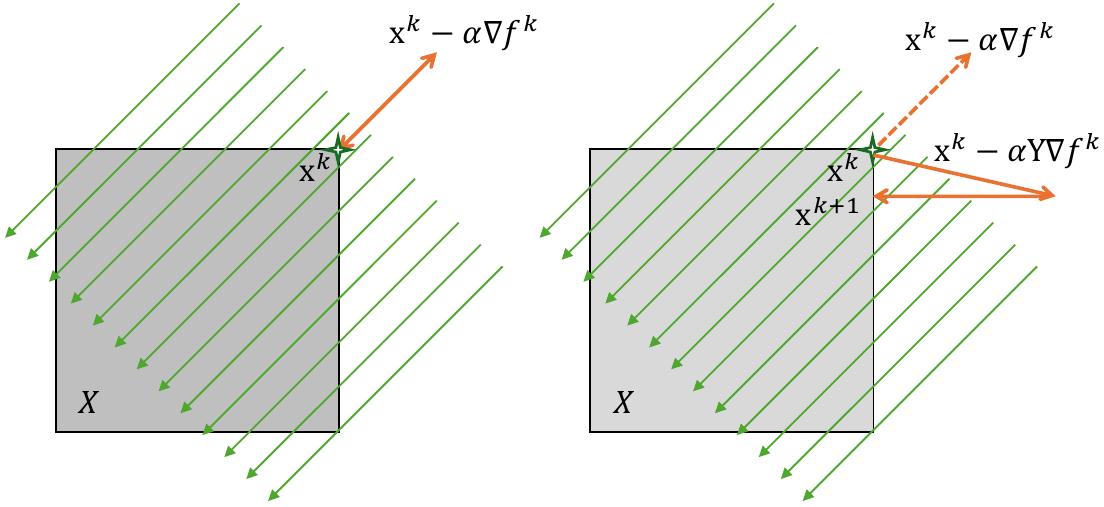}
  \caption{An illustration showing that the two-metric approach is not a descent iteration, where $\mathbf{x}^k$ is the current iterate which is also the optimum, $X$ is the feasible region, and $\nabla f^k$ is the gradient at $\mathbf{x}^k$. With (plain) gradient projection on the left panel, the iterate stays at the optimum, i.e. $\mathbf{x}^{k+1}=\mathbf{x}^{k}$. With the two-metric approach on the right panel where $\mathbf{Y}$ represents the gradient scaling metric, $\mathbf{x}^{k+1}$ leaves the optimum.}
  \label{fig:two-metric}
\end{figure}

Computing (\ref{grad:two-metric}) for node $i$ now requires only information exchange with its physical neighbours. However, the two-metric approach does not necessarily lead to algorithm convergence since it is not in general a descent iteration \cite{Bertsekas1999}. This is illustrated in Fig. \ref{fig:two-metric} with an example. For further developments, we define 
\begin{align} \label{eq:q_proj}
 & \mathbf{\dot{q}}^{k+1} = \\
 &  \quad \left[ q_i^k - \alpha [ [\mathbf{X}^{-1} \mathbf{C} \mathbf{q}^k]_i + \lambda_i^{k+1}-\mu_i^{k+1}+ r^p q_i^k], i \in \mathcal{N}^+ \right], \nonumber
\end{align}
which are tentative but not necessarily feasible setpoints to be projected for all nodes, and define 
\begin{align} \label{eq:X}
  \mathcal{X}^k = [\underline{q}_1^k, \overline{q}_1^k]\times[\underline{q}_2^k, \overline{q}_2^k]\times \cdots \times [\underline{q}_{N}^k, \overline{q}_{N}^k],
\end{align}
where $\times$ is the Cartesian product of sets.

\subsection{Proposed distributed approach}\label{sec:proposed}
It has been shown in \cite{Bertsekas1999} that, to ensure convergence, the projection should use the distance measured in terms of the norm $\lVert \mathbf{z} \rVert _\mathbf{X} = \sqrt{\mathbf{z}^\intercal \mathbf{X} \mathbf{z}} $ due to the scaling matrix $\mathbf{X}^{-1}$ \cite{Bertsekas1999}, that is:
\begin{equation}\label{grad:consistent}
    \mathbf{q}^{k+1} = \underset{\mathbf{u}\in\mathcal{X}^k}{\operatorname{argmin}} \quad \frac{1}{2} (\mathbf{u}-\mathbf{\dot{q}}^{k+1})^\intercal \mathbf{X} (\mathbf{u}-\mathbf{\dot{q}}^{k+1}).
   \end{equation}

This however contradicts the distributed communication architecture of the overall OFO algorithm since, unlike the Euclidean projection, solving (\ref{grad:consistent}) again requires global information. As a solution, \cite{Qu2020} proposed dualization of the DER capacity limits, combined with the Euclidean projection, to ensure the DER constraint satisfaction. However, this approach requires an excessively large number of iterations to reach convergence, significantly hindering its performance in online implementation.

  To tackle this challenge, this paper proposes an iterative projected gradient descent (PGD) algorithm to solve (\ref{grad:consistent}) that exchanges information only between physical neighbours in a backward-forward manner. This leads to our nested feedback optimization approach, where the outer loop comprises the OFO iterations giving tentative reactive power setpoints, and the inner loop solves the non-Euclidean projection problem (\ref{grad:consistent}) to generate feasible setpoints.

  Given a $\mathbf{u}^{k,\tau} \in \mathcal{X}^k$ where $\tau = 0, 1, 2, \cdots, T$ is the index for the inner loop, the PGD algorithm proceeds to compute $\mathbf{u}^{k,\tau+1} \in \mathcal{X}^k$ using (\ref{eq:proposed}) where $ \mathbf{X} (
    \mathbf{u}^{k,\tau}-\mathbf{\dot{q}}^{k+1}
  )$ is the gradient of the objective function of (\ref{grad:consistent}) at $\mathbf{u}^{k,\tau}$.
  \begin{equation} \label{eq:proposed}
      \mathbf{u}^{k,\tau+1} = [\mathbf{u}^{k,\tau} - \alpha^{u} \mathbf{X} (
          \mathbf{u}^{k,\tau}-\mathbf{\dot{q}}^{k+1}
        )]_{\mathbf{\underline{q}}^k}^{\mathbf{\overline{q}}^k}.
    \end{equation}

  The remaining challenge is to compute the gradient in a distributed manner. To achieve this, we exploit the underlying tree topology of radial distribution networks, which enables information propogation in a backward-forward manner. In \cite{Su2023}, this procedure is derived using LDU factorization. While backward-forward strategies have been used on tree graphs, our contribution lies in leveraging them to solve the non-Euclidean projection problem, completing a distributed voltage control scheme that optimally coordinates DERs on a fast timescale, operates in a feedback-based manner, and preserves end-users' privacy by keeping voltage data local. For notational simplicity, we denote $\mathbf{u}^{k,\tau}-\mathbf{\dot{q}}^{k+1}$ as $\boldsymbol{\omega}^{k,\tau}$ and denote its $i$-th element as $\omega_i^{k,\tau}$, which is locally known by node $i$.

    In the backward step, for each node $i$, upon collecting $\xi_j^{k,\tau}$ for all $j \in \mathcal{C}_i$ where  $\mathcal{C}_i$ represents the set of child nodes of $i$, it sends to its unique parent node 
\begin{equation}\label{eq:backward}
  \xi_i^{k,\tau} = \sum_{j\in\mathcal{C}_i} \xi_j^{k,\tau} + \omega_i^{k,\tau}.
\end{equation}
Note that this step runs recursively from the farthest nodes up to but not including the closest nodes to the point of common coupling (PCC), where $\xi_j^{k,\tau}$ is recursively defined. In a nutshell, $\xi_i^{k,\tau}$ represents the sum of $\omega_n^{k,\tau}$ where $n$ belongs to all of its downstream nodes including itself.

    In the forward step, for each node $i$, upon receiving $\zeta_h^{k,\tau}$ from its unique parent node, it sends to its child node(s) 
    \begin{equation}\label{eq:forward}   
      \zeta_i^{k,\tau} = \zeta_h^{k,\tau} + x_{hi} \xi _{i}^{k,\tau}.
  \end{equation}
  Similarly, this step runs recursively from the closest nodes  to the PCC up to but not including the farthest nodes, where $\zeta_h^{k,\tau}$ is recursively defined. Likewise, $\zeta_i^{k,\tau}$ represents the sum of $x_{mn} \xi _{n}^{k,\tau }$ where $n$ belongs to all of its upstream nodes including itself.

Based on the information received from its child node(s) and parent node, each node $i$ can then determines its entry in $\mathbf{X}\boldsymbol{\omega}^{k, \tau}$ as
\begin{equation}\label{eq:result} 
  [\mathbf{X}\boldsymbol{\omega}^{k, \tau}]_i = \zeta_h^{k,\tau} + x_{hi} \xi _{i}^{k,\tau }.
\end{equation}
Note that in this process, each node only communicates with its physical neighbours once. The overall nested feedback optimization algorithm is summarized in Algorithm \ref{alg:nofo}.

\begin{algorithm}[t]
  \caption{Nested feedback optimization algorithm}\label{alg:nofo}
  \KwData{$\mathbf{q}^0$, $\boldsymbol{\lambda}^0$, $\boldsymbol{\mu}^0$}
  \KwResult{$\mathbf{q}^k$,  $k\geq 0$}
  Set k = 0\;
  \While{$k\geq 0$}{
  Implement $\mathbf{q}^k$ and collect voltage measurement $\tilde{\mathbf{v}}(\mathbf{q}^k)$\;
  Update $\boldsymbol{\lambda}^{k+1}$ and $\boldsymbol{\mu}^{k+1}$ via (\ref{grad:lambda})-(\ref{grad:mu})\;
  Communicate $\mathbf{q}^k$ between physical neighbours and calculate $\dot{\mathbf{q}}^{k+1}$ via (\ref{eq:q_proj})\;
  Set $\tau=0$\;
  \While{$\tau < T$}{Use the backward-forward algorithm (\ref{eq:backward})-(\ref{eq:result}) to compute  $ \mathbf{X} (\mathbf{u}^{k,\tau}-\mathbf{\dot{q}}^{k+1} )$
    (choose $\mathbf{u}^{k,0}$ as $\mathbf{q}^k$)\;
    Calculate next $\mathbf{u}^{\tau+1}$ via (\ref{eq:proposed})\;
   Set $\tau = \tau +1$\;
  }
  Set $\mathbf{q}^{k+1} = \mathbf{u}^T$\;
  Set $k = k+1$. 
  }
  \end{algorithm}

  \begin{remark}
    In principle, the same approach could be applied to (\ref{grad:centralized}) to develop a distributed strategy. However, this would require sharing dual variables. Due to the correlation between voltage and active power, exposing these variables could enable external entities to infer sensitive load information, which is typically considered private. In contrast, our strategy avoids sharing dual variables.
  \end{remark}

  \begin{remark}
The proposed method assumes that each DER has access to topological information, such as knowledge of its immediate neighbors. This information is assumed to be obtained during the algorithm's planning phase. However, we acknowledge that this may not always be feasible. Therefore, exploring approaches that reduce or eliminate this dependency represents an interesting research direction to enhance the practicality and adaptability of the algorithm.
  \end{remark}

    \begin{remark}
The addition or removal of DERs may require reconfiguration of the communication graph between agents. However, if each node is interpreted as a virtual DER—potentially representing an aggregation point in the network—the communication structure can remain unchanged. Under this setting, the presence or absence of individual physical DERs does not necessarily impact the agent-level communication.
    \end{remark}
  



\subsection{Convergence analysis}
The convergence analysis of the nested algorithm is based on the following two assumptions.

\begin{assumption}
  The linear relation (\ref{lpf:vcomp}) holds between voltages and reactive power injections.
\end{assumption}


\begin{assumption}
  The inner iterations reach convergence after $T$ iterations.
\end{assumption}

The inner loop represents a standard Euclidean gradient projection algorithm to solve the quadratic program (\ref{grad:consistent}). Note that the backward-forward step is only used to gather gradient information; each agent then performs a projected gradient descent step. Consequently, \{$\mathbf{u}^\tau$\} is a converging sequence by invoking Proposition 2.3.2 in \cite{Bertsekas1999} with $0<\alpha^u<2/\lambda_{max}(\mathbf{X})$ where $\lambda_{max}(\mathbf{X})$ is the largest eigenvalue of $\mathbf{X}$. Since we cannot choose $T$ to be arbitrarily large in a practical system, this algorithm will generate an approximate solution to (\ref{grad:consistent}). When $T$ is chosen sufficiently large, the algorithm will converge arbitrarily close to the optimum of (\ref{grad:consistent}) and the approximation error will be negligible. In Section \ref{sec:sim}, we demonstrate its convergence with numerical results.

Under the two assumptions, the overall nested algorithm represents a scaled PDGP algorithm to solve (\ref{opt}). By a transformation of variables defined by: $\mathbf{q}' = \mathbf{X}^{\frac{1}{2}} \mathbf{q}$ and $\mathbf{v}' = \gamma \mathbf{v}$ where $\gamma=\sqrt{\frac{\alpha^d}{\alpha}}$ is the squared root of the ratio between the dual and primal step sizes, the problem (\ref{opt}) can be cast as an equivalent quadratic program in the space of $\mathbf{q}'$ and $\mathbf{v}'$ as in (\ref{opt:transformed}).\looseness=-1
\begin{subequations}\label{opt:transformed}
  \begin{align}
    \underset{\mathbf{q'}}{\operatorname{minimize }} & \quad \frac{1}{2} \mathbf{q'}^\intercal \mathbf{X}^{-\frac{1}{2}} \mathbf{C} \mathbf{X}^{-\frac{1}{2}} \mathbf{q'} , \label{trans:obj}\\
    \operatorname{s.t. } & \quad \gamma \mathbf{\underline{v}} \leq \mathbf{v'} \leq \gamma \mathbf{\overline{v}}: \boldsymbol{\mu'},\boldsymbol{\lambda'} ,\label{trans:cons:voltage}\\
    & \mathbf{\underline{q}} \leq  \mathbf{X}^{-\frac{1}{2}} \mathbf{q'}  \leq \mathbf{\overline{q}}. \label{trans:cons:reactive}
  \end{align}
\end{subequations}

The unscaled regularized PDGP algorithm to solve (\ref{opt:transformed}) includes the following iterations:
\begin{subequations}\label{pdgp:transformed}
\begin{align}
  &\boldsymbol{\lambda'}^{k+1} = [\boldsymbol{\lambda'}^{k} + \alpha (\gamma \mathbf{\tilde{v}}-\gamma \mathbf{\overline{v}}-r^d \boldsymbol{\lambda'}^{k})]^+,\\
  &\boldsymbol{\mu'}^{k+1} = [\boldsymbol{\mu'}^{k} + \alpha (\gamma \mathbf{\underline{v}}-\gamma \mathbf{\tilde{v}}-r^d \boldsymbol{\mu'}^{k})]^+,\\
  & \mathbf{q'}^{k+1} = \underset{\mathbf{q',\mathbf{\underline{q}} \leq  \mathbf{X}^{-\frac{1}{2}}\mathbf{q'}  \leq \mathbf{\overline{q}}}}{\operatorname{argmin}} \lVert \mathbf{q'}-\big[\mathbf{q'}^{k} - \alpha [\mathbf{X}^{-\frac{1}{2}}    \\
  &\qquad \cdot \mathbf{C} \mathbf{X}^{-\frac{1}{2}} \mathbf{q'}^k +  \gamma \mathbf{X}^{\frac{1}{2}} (\boldsymbol{\lambda'}^{k+1}-\boldsymbol{\mu'}^{k+1}) + r^p \mathbf{q'}^k] \big] \rVert_2^2. \nonumber
\end{align}
\end{subequations}
which are equivalent to (\ref{grad:lambda})-(\ref{grad:mu}) and (\ref{eq:q_proj})-(\ref{grad:consistent}) \cite[Eqs. (2.37) and (2.38)]{Bertsekas1999}. Invoking Theorem 3 in \cite{Bernstein2019} with $\alpha$ chosen using \cite[Eq. (34)]{Bernstein2019}, (\ref{pdgp:transformed}) then converges to the optimizer of the regularized saddle-point problem of (\ref{opt:transformed}). Consequently, our proposed nested algorithm also converges to the optimizer of the regularized saddle-point problem of (\ref{opt}). 

\begin{remark}
While the linear voltage-reactive power relation is assumed to synthesize the controller and analyze its convergence, AC power flow is run in the simulation to calculate the actual voltages. Such an assumption was also made in the proofs of \cite{Bolognani2019,Tang2024}. 
\end{remark}

\section{Case study}\label{sec:sim}
\subsection{Case description}
\begin{figure}[t]
    \centering
    \includegraphics[width=\linewidth]{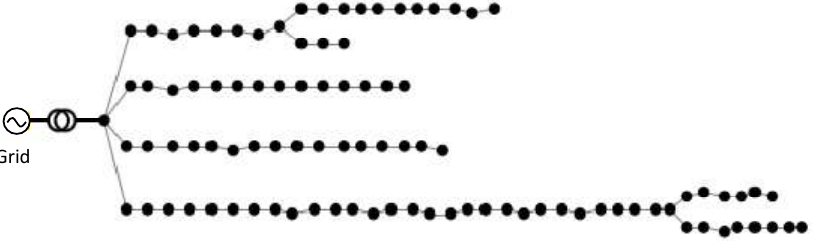}
    \caption{A 96-bus test system from Simbench \cite{Meinecke2020}.}
    \label{fig:grid}
  \end{figure}

\begin{figure}[t]
    \centering
    \subfloat[Trajectories of the cost function]{\includegraphics[width=\columnwidth]{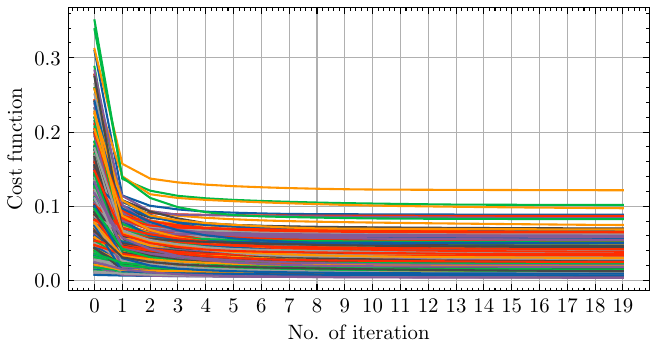}}\\
    \subfloat[Distribution of the relative cost function (w.r.t. the optimum)]{\includegraphics[width=\columnwidth]{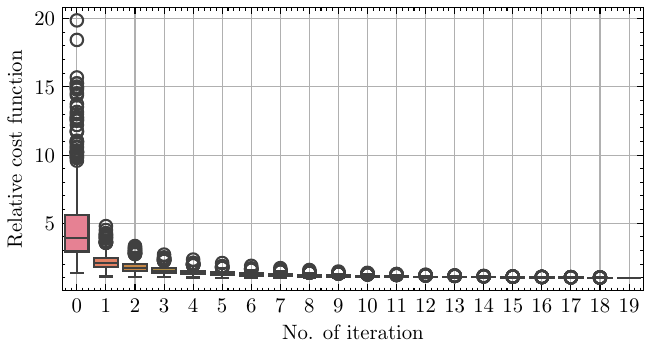}}
    \caption{Results of 1000 random problem instances.}
    \label{fig:converge}
  \end{figure}

In the numerical simulation, we study a 100\% penetration scenario of photovoltaics (PVs) in a 96-bus system, i.e. we assume every node has a PV installation. The test system is adopted from Simbench \cite{Meinecke2020} and is shown in Fig. \ref{fig:grid}, which is expected to experience overvoltage issues. The network topology and impedance data are the same as those in the original dataset. For each bus, a PV installation with a randomly generated DC capacity from 3-10 kW is assumed. The PV inverters are assumed to be oversized by 20\% to provide sufficient reactive power. For the dynamic case study, generation and load data with a 6-second resolution are used. The PV profiles are linearly interpolated using the \mbox{\textit{HelioClim-3}} dataset \cite{helioclim} with a 1-minute resolution. Base load profiles are obtained by aggregating 1-second resolution data from the \textit{ECO} dataset \cite{beckel2014nilm}. AC power flow problems are solved with a high-performance library \textit{PowerGridModel} \cite{Xiang2023}.


The reactive power cost matrix $\mathbf{C}$ is chosen as the identity matrix. The lower and upper voltage limits are 0.95 and 1.05 pu, respectively. The unit for power is kW/kVar. For the inner loop, $\alpha^u$ is chosen as $0.99 \cdot 2/\lambda_{max}(\mathbf{X})$. For the outer loop, the step sizes are chosen with a trial-and-error strategy \cite{Qu2020} as $\alpha^d=10^6$ and $\alpha=3\times10^{-4}$. Following the guideline in \textit{Remark} \ref{re:reg}, the regularization parameters 
$r_p$ and $r_d$ are set to $10^{-4}$. The algorithm performs well with such small regularization, likely due to the strongly convex objective function in (\ref{opt}). For each generation and load data point, 3 OFO iterations are run which yields tentative reactive power setpoints, followed by 10 inner iterations, i.e. we assume that each inner iteration takes 200 milliseconds. It is crucial that the system responds faster than the DER setpoint update rate. In our application, setpoints are updated on a (sub-)second timescale, and the system is assumed to react even more quickly. Similarly, \cite{multiarea,DallAnese2018,Bernstein2019} consider DER update time of 0.1, 0.33, and 1 second, respectively. Another relevant example is droop control, where voltage measurements and setpoint updates occur in real time, typically within the subsecond range \cite{Baker2017}. Finally, the nested algorithm requires only basic arithmetic operations and is computationally very efficient. For a series implementation of a 4-hour simulation, i.e. in total 7200 iterations, it takes 6.7 seconds, averaging 0.93 milliseconds per iteration.


\begin{figure}[t]
    \centering
    \includegraphics[width=\linewidth]{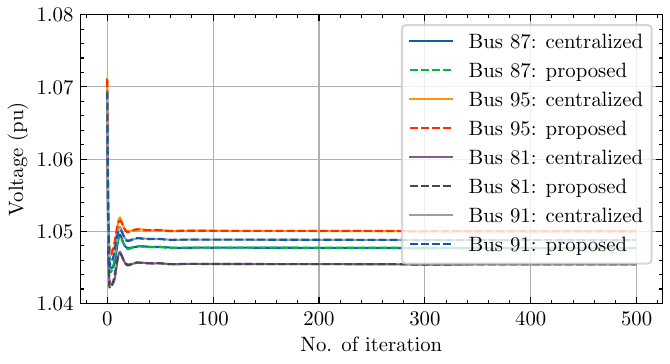}
    \caption{Voltage profiles over 500 iterations for several buses in the static case using the centralized and the proposed approaches. We count both the outer and inner iterations for our proposed approach.}
    \label{fig:voltage}
\end{figure}

\begin{figure}[t]
    \centering
    \includegraphics[width=\linewidth]{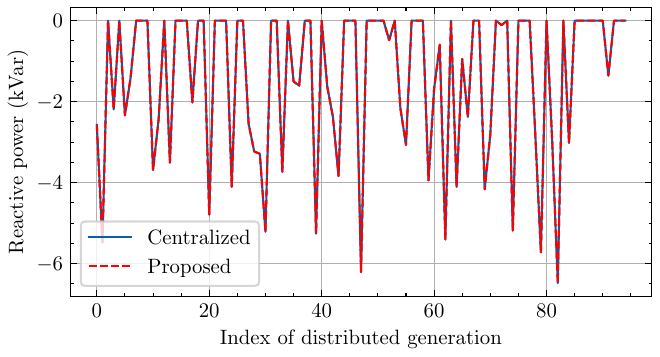}
    \caption{Reactive power profiles after 500 iterations for the centralized and the proposed approaches.}
    \label{fig:reactive}
\end{figure}

\begin{figure}[t]
  \centering
\includegraphics[width=\linewidth]{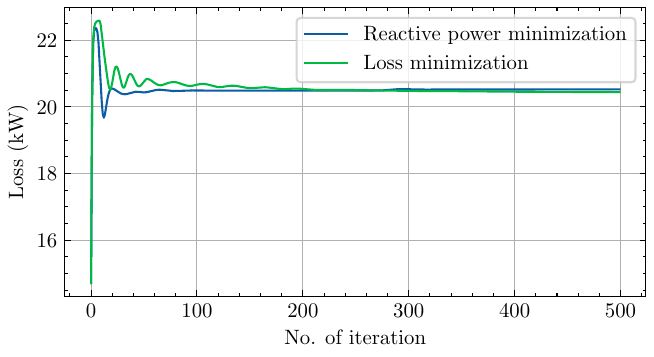}
  \caption{Loss comparison between the objectives of reactive power minimization and loss minimization.}
  \label{fig:loss}
\end{figure}

\begin{figure*}[t]
    \centering
    \subfloat[No control]{\includegraphics[width=\columnwidth]{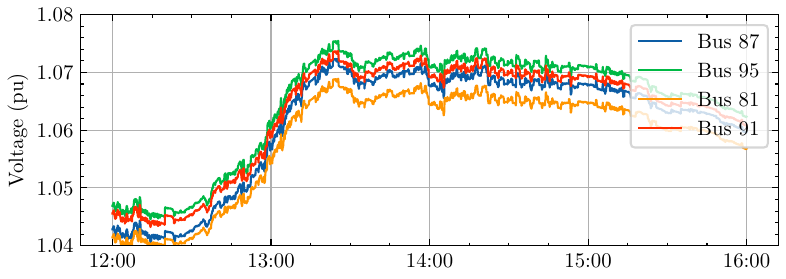}}
    \subfloat[Centralized]{\includegraphics[width=\columnwidth]{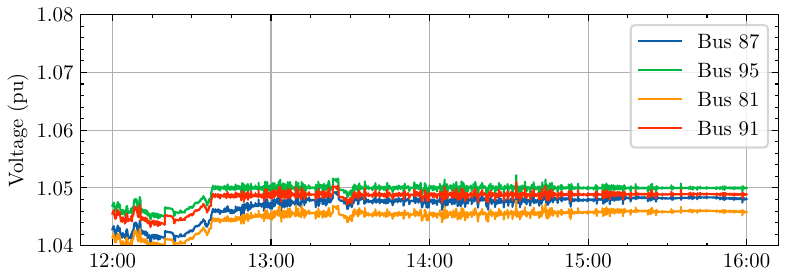}}\\\vspace{-10pt}
    \subfloat[Proposed]{\includegraphics[width=\columnwidth]{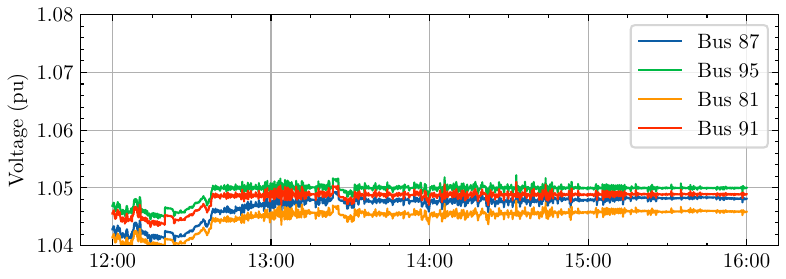}}
    \subfloat[Two metric]{\includegraphics[width=\columnwidth]{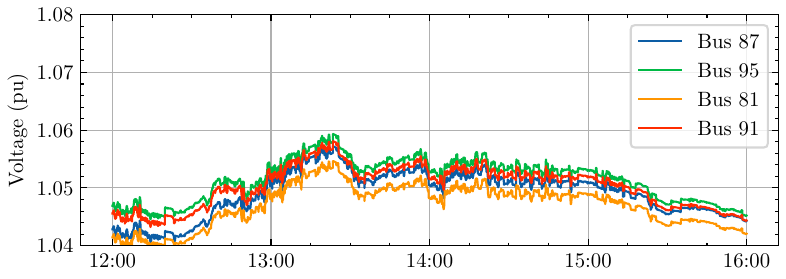}}\\\vspace{-10pt}
    \subfloat[Tang 2024 \cite{Tang2024}]{\includegraphics[width=\columnwidth]{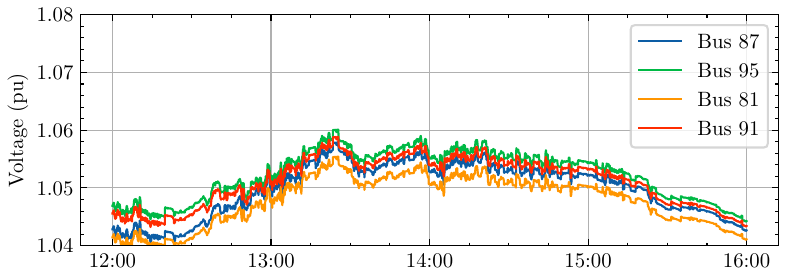}}
    \subfloat[Qu 2020 \cite{Qu2020}]{\includegraphics[width=\columnwidth]{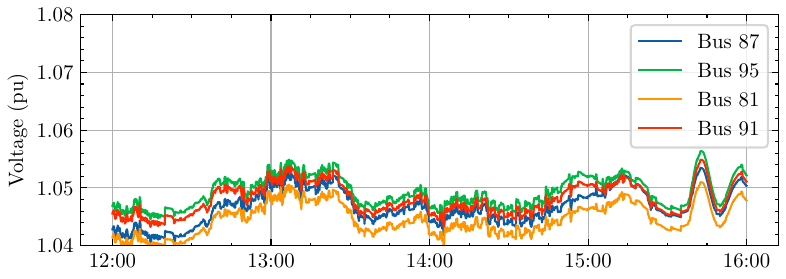}}\\\vspace{-10pt}
    \subfloat[Droop control \cite{Vergara2020,Jahangiri2013}]{\includegraphics[width=\columnwidth]{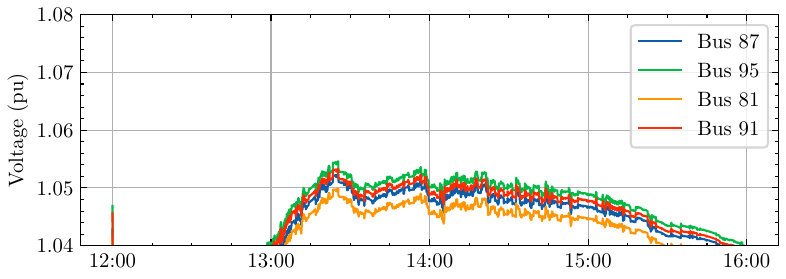}}
    \subfloat[Summary for bus 95]{\includegraphics[width=\columnwidth]{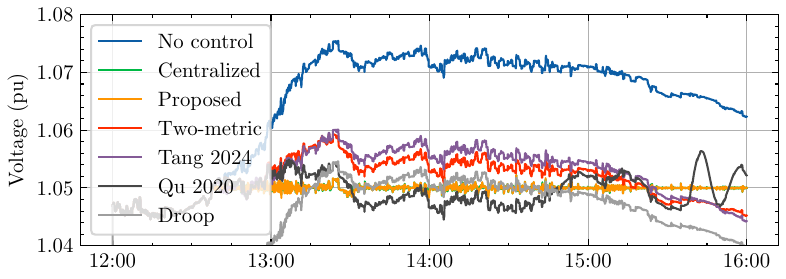}}
    \caption{Voltage profiles under different approaches in the dynamic simulation.}
    \label{fig:voltage_dynamic}
  \end{figure*}

  \begin{figure*}[t]
    \centering
    \subfloat[Centralized]{\includegraphics[width=\columnwidth]{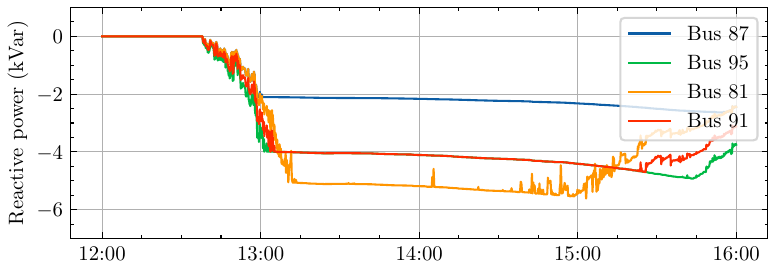}}
    \subfloat[Proposed]{\includegraphics[width=\columnwidth]{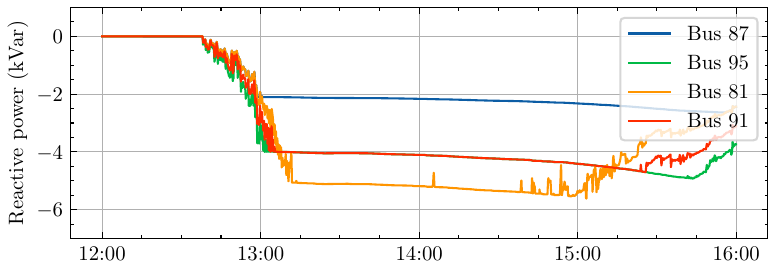}}\\\vspace{-10pt}
    \subfloat[Two metric]{\includegraphics[width=\columnwidth]{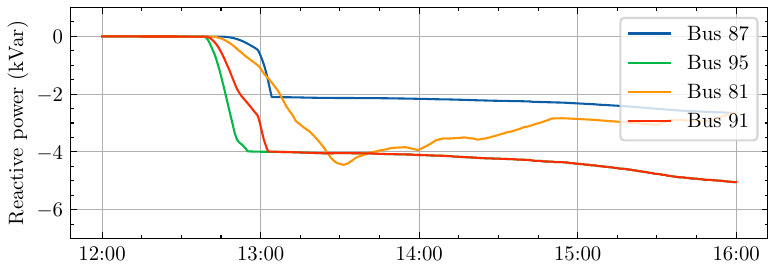}}
    \subfloat[Tang 2024 \cite{Tang2024}]{\includegraphics[width=\columnwidth]{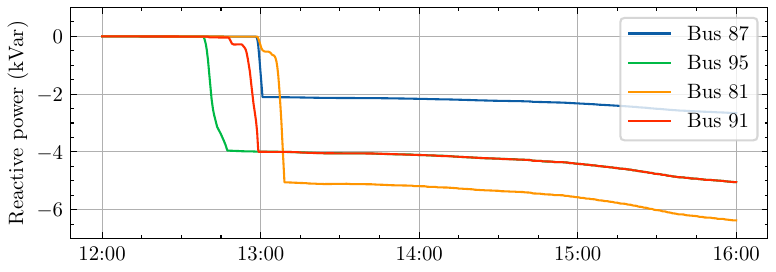}}\\\vspace{-10pt}
    \subfloat[Qu 2020 \cite{Qu2020}]{\includegraphics[width=\columnwidth]{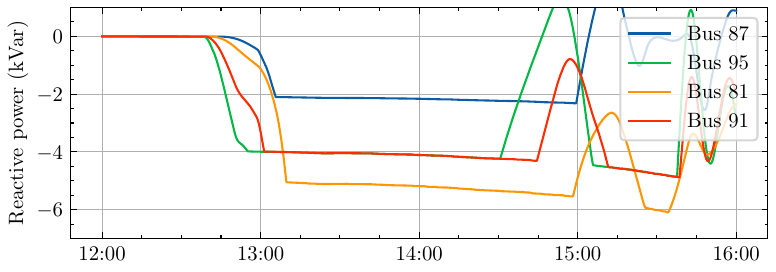}}
    \subfloat[Droop control \cite{Vergara2020,Jahangiri2013}]{\includegraphics[width=\columnwidth]{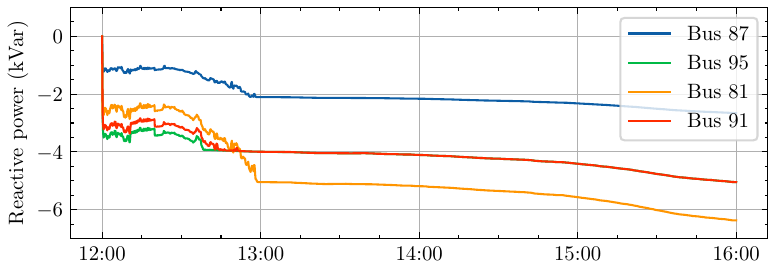}}
    \caption{Reactive power profiles under different approaches in the dynamic simulation.}
    \label{fig:power_dynamic}
  \end{figure*}

  \subsection{Convergence of inner loop}

To demonstrate the convergence of the inner loop, 1000 random problem instances for (\ref{grad:consistent}) are generated. Figure \ref{fig:converge}a illustrates that the cost function drops significantly within a few iterations, leading to rapid algorithm convergence. Additionally, Fig. \ref{fig:converge}b depicts the distribution of the relative cost function with respect to the optimum. After approximately 10 iterations, the cost function stabilizes. Thus, in this study, we set the inner loop to run for 10 iterations, ensuring a balance between solution accuracy of the inner loop and the update frequency of the outer loop.

\subsection{Static case results}\label{sec:result-static}
In the static case, the generation and load data are kept unchanged to study the convergence of the nested approach. The simulation uses data at 12:00. Figure \ref{fig:voltage} shows that the proposed approach suppresses the initial voltage limit violations quickly and converges after around 60 outer iterations. The approach also utilizes the available voltage limit as efficiently as the centralized one. Figure \ref{fig:reactive} compares the converged reactive power setpoints, which shows that the centralized and the proposed approaches converge to the same solution. This is reasonable since the gradient scaling only alters the converging trajectory, but does not change the solution to (\ref{opt}). Finally, \mbox{Fig. \ref{fig:loss}} demonstrates that under overvoltage issues caused by excessive PV generation, the chosen objective function of minimizing reactive power usage aligns with minimizing system losses. However, under undervoltage conditions, the outcome may differ, as increased reactive power injection could potentially help lower system losses. This highlights a tradeoff between the distributed communication architecture and the flexibility in selecting optimization objectives.

\subsection{Dynamic case results}
In the dynamic case, we study the performance of the approaches in a time-varying environment, where the approaches are implemented online to track the time-varying optimizer of (\ref{opt}) without waiting for them to converge. \mbox{Figure \ref{fig:voltage_dynamic}}a shows significant voltage limit violations without control. Figures \ref{fig:voltage_dynamic}b and \ref{fig:voltage_dynamic}c demonstrate that the centralized and our proposed distributed nested feedback optimization approaches both successfully enforce the voltage limit, although temporary voltage limit violations are inevitable due to the corrective nature of these algorithms. The simple two-metric approach and the heuristic approach in \cite{Tang2024} result in prolonged voltage limit violations as seen in Figs. \ref{fig:voltage_dynamic}d and \ref{fig:voltage_dynamic}e, respectively. Figure \ref{fig:voltage_dynamic}f shows that the distributed approach in \cite{Qu2020} does not track the time-varying optimizers well and experiences oscillations at the end of the simulation period. The discrepancy between the actual setpoints and the algorithm iterates (without projection) potentially renders the algorithm slow and unstable. \mbox{Figure \ref{fig:voltage_dynamic}g} shows that the local droop control approach does not enforce the voltage limit due to lack of coordination. Finally, for clarity, Fig. \ref{fig:voltage_dynamic}h presents voltages for the most remote, and therefore, the most sensitive bus (bus 95) under different approaches.\looseness=-1

Furthermore, Fig. \ref{fig:power_dynamic} shows reactive power profiles under different approaches. Shown in Figs. \ref{fig:power_dynamic}a and \ref{fig:power_dynamic}b, our proposed distributed approach achieves similar profiles with the centralized one, with an average setpoint deviation of \mbox{0.01 kVar}. The two-metric approach and the heuristic approach do not work well in this case study. The approach in \cite{Qu2020} experiences oscillations. The local droop control approach does not in theory or in practice track the optimizer of (\ref{opt}).

Finally, we define the average voltage violation (AVV) in (\ref{eq:avv_dist}), where $k$ is the iteration index and $K$ is the total number of iterations.
\begin{align} \label{eq:avv_dist}
    \operatorname{AVV}=\frac{1}{K}  \sum_{k=1}^{K} \left([\tilde{v}^k-\overline{v}]^+ + [\underline{v}-\tilde{v}^k]^+ \right).
\end{align}

\begin{table}[t]
    \centering
    \caption{Average voltage violations and losses for various approaches.}
    \begin{tabular}{c|c|c|c} 
     \hline
     Approach & AVV (pu) & AVV w.r.t. central. & \text{Loss (kWh)}\\
     \hline
     No control & $1.5\times10^{-2}$ & 192x &\text{48.2}\\
     Centralized &  $8.0\times10^{-5}$ &1x&\text{64.0}\\
     Proposed &  $9.9\times10^{-5}$& 1.2x &\text{64.0}\\
     Two-metric &  $2.5\times10^{-3}$ & 31x &\text{59.3}\\
     Tang 2024 \cite{Tang2024} &  $3.3\times10^{-3}$ & 41x &\text{57.4}\\
     Qu 2020 \cite{Qu2020} &  $8.6\times10^{-4}$ &11x&\text{65.9}\\
     Droop control \cite{Vergara2020}& $4.5\times10^{-4}$&6x&\text{67.8}\\
     \hline
    \end{tabular}
    \label{table:avv}
    \end{table}
Table \ref{table:avv} shows the AVV values for bus 95 (the most sensitive bus) and system losses under various approaches. An AVV of $9.9\times10^{-5}$ pu is achieved for the proposed approach, which is slightly higher than that of the centralized approach. The latter however requires a centralized communication architecture and is thus susceptible to failure of the central coordination unit. The proposed nested feedback approach significantly outperforms other existing distributed or local approaches. Because reactive power is absorbed to mitigate voltage rise, system losses increase. The proposed approach results in the same losses as the centralized method.

\subsection{Impact of measurement noise}

\begin{figure}[t]
  \centering
\includegraphics[width=\linewidth]{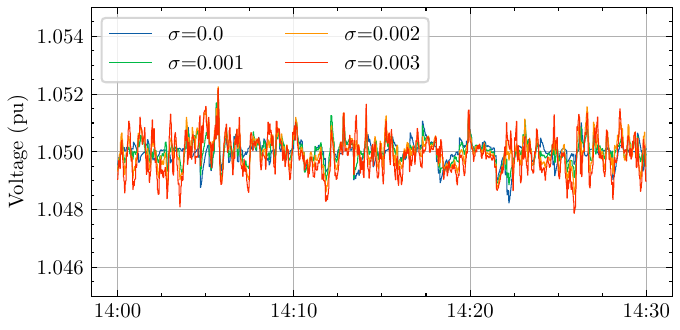}
  \caption{Voltages for bus 95 under measurement noise.}
  \label{fig:noise}
\end{figure}

This section examines the impact of measurement noise on the voltage regulation performance. The noise is modeled using a zero-mean Gaussian distribution with a standard deviation of $\sigma$. Figure \ref{fig:noise} illustrates its effect. As measurement noise increases, voltage fluctuations become more pronounced. Nevertheless, the controller successfully maintains voltages within safe limits. The temporary voltage limit violations can be mitigated by setting a tighter limit, e.g. 1.048 pu.

\subsection{Impact of node failures}

\begin{figure}[t]
  \centering
\includegraphics[width=\linewidth]{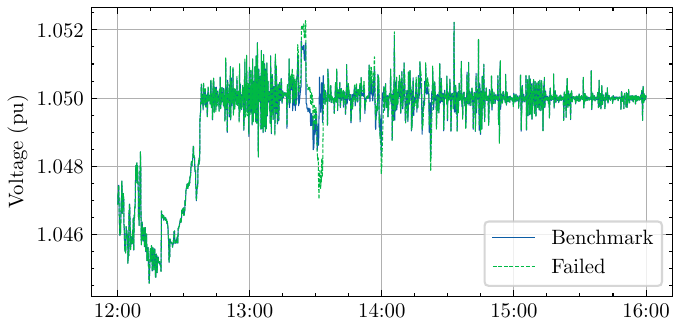}
  \caption{Voltages for bus 95 under a failed node.}
  \label{fig:fail}
\end{figure}

\begin{figure}[t]
  \centering
\includegraphics[width=\linewidth]{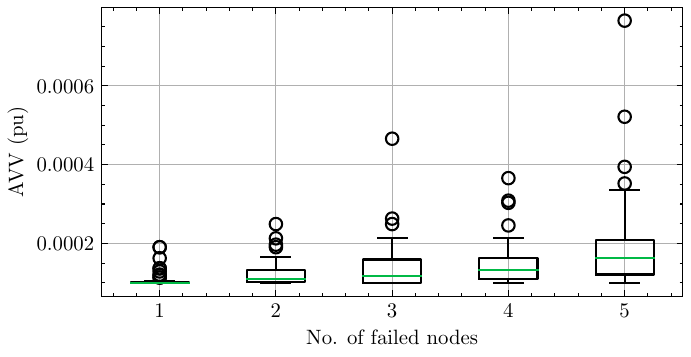}
  \caption{Impact of node failres on AVV.}
  \label{fig:fail2}
\end{figure}

  In this section, we demonstrate the robustness of the proposed strategy against random node failures. When a node fails, its neighbouring nodes are assumed to bypass it and continue communication. Figure \ref{fig:fail} illustrates an example voltage profile with a failed node, showing that the system maintains the voltage limit despite a random node failure. Figure \ref{fig:fail2} depicts how AVV increases with the number of failed nodes but remains within a reasonable range, indicating the algorithm's robustness. Unlike centralized approaches, where failure of the central coordination unit leads to a complete loss of voltage regulation, the proposed method continues to function effectively. Enhancing system resilience through the development of failure detection and bypass mechanisms is an important direction for future research.

\subsection{Impact of update frequency}

\begin{figure}[t]
  \centering
\includegraphics[width=\linewidth]{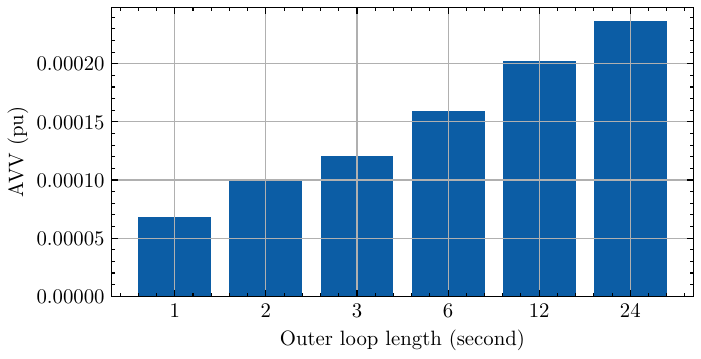}
  \caption{AVV under various update frequency.}
  \label{fig:frequency}
\end{figure}

  Figure \ref{fig:frequency} quantifies the impact of the OFO controller update frequency on its voltage regulation performance. Note that the simulated generation and load data have a resolution of 6 seconds. As expected, more frequent updates improve performance. However, even with an outer iteration length of 24 seconds, the performance deterioration remains limited. This suggests that the method retains effectiveness even under slower update rates. While we assume relatively fast update in the inner loop (e.g. 200 milliseconds per iteration), we acknowledge that this may be optimistic in large or deep networks. Therefore, this section serves to assess the controller's robustness to slower updates, which relaxes the need for low-latency communication. The optimal update frequency depends on the communication infrastructure and requires a cost-effectiveness analysis. Future work will focus on analyzing the scalability of the controller under more realistic communication constraints, including increased latencies and asynchronous operation.

\begin{figure}[t]
  \centering
\includegraphics[width=\linewidth]{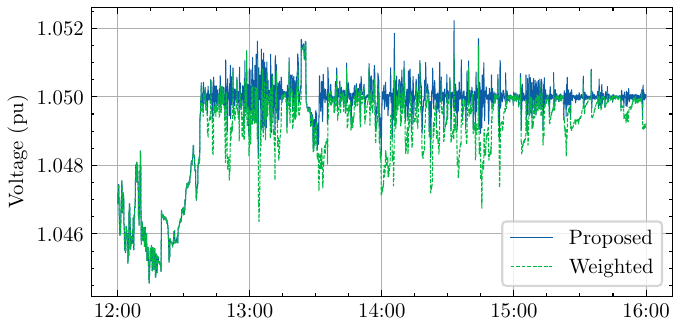}
  \caption{Voltages for bus 95 under various objective functions.}
  \label{fig:bolog}
\end{figure}

\subsection{Impact of the objective function}

In this section, we compare our objective function with the alternative objective of minimizing $\mathbf{q}^\intercal \mathbf{X} \mathbf{q}$, as adopted in \cite{Bolognani2019,Ortmann2020_dist,Tang2021}. Figure \ref{fig:bolog} shows that this weighted objective function (i.e. weighted by $\mathbf{X}$) also
effectively enforces the voltage limit in the dynamic case. The voltage profile obtained using the weighted cost function yields a lower average voltage violation (AVV) of $3.0\times10^{-5}$ pu but exhibits greater variation. Given that the system faces overvoltage issues, the voltage constraints are expected to be binding at the optimal solution. Accordingly, we use 1.05 pu as the reference voltage for the most remote bus. Under this setting, the voltage profile from the weighted cost function results in an average deviation of $6.3\times10^{-4}$ pu, which is significantly higher than that achieved with our proposed formulation ($2.3\times10^{-4}$ pu). Moreover, the weighted formulation assigns higher costs to PV inverters located at the end of distribution feeders, leading to suboptimal and unfair utilization of available reactive power capacity. Consequently, it results in a total reactive power use of 460.4 kVarh, compared to 390.9 kVarh in our formulation. Reducing reactive power absorption can be beneficial for lowering system losses and extending PV inverter lifetime. In this simulation, the weighted cost function yields a loss of 65.4 kWh, which is higher than the 64.0 kWh loss achieved with our formulation. Finally, our formulation is flexible and can accommodate this alternative objective function if considered beneficial.

\subsection{Active power control}

\begin{figure}[t]
  \centering
\includegraphics[width=\linewidth]{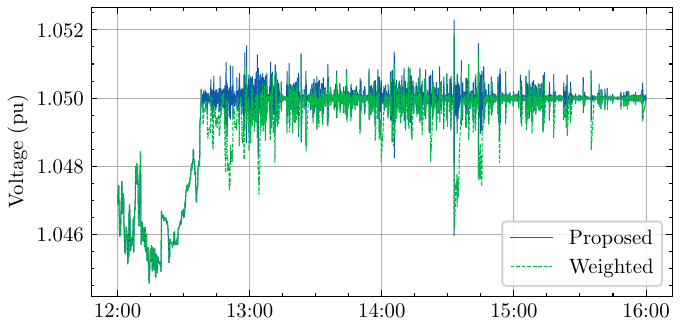}
  \caption{Voltages for bus 95 under various objective functions with active power control.}
  \label{fig:bolog_p}
\end{figure}

To further demonstrate the significance of the proposed formulation, we apply the algorithm for active power control to mitigate voltage limit violations in scenarios where reactive power capacity is insufficient. As shown in Fig. \ref{fig:bolog_p}, both formulations successfully enforce the voltage limit. However, the weighted formulation leads to 17.5\% active power curtailment (1649.7 kWh generated out of a possible 2000.7 kWh), whereas the proposed formulation reduces curtailment to only 7.7\% (1847.2 kWh generated). This highlights the practical advantage of the proposed approach.

\section{Conclusion}\label{sec:concl}

In this paper, we focused on the distributed voltage regulation problem in distribution systems and proposed a nested feedback optimization approach. We theoretically analyzed its convergence. Our simulation results showed that the approach, while only requiring short-range communication between physical neighbours, achieved satisfactory voltage regulation and outperformed existing distributed and local approaches. The proposed distributed approach does not extend to congestion management while the centralized does so.  Extending the approach to unbalanced systems is an interesting future direction, where the focus is to construct a positive definite $\mathbf{X}$ with a sparse inverse matrix. Finally, the approach can be extended to joint active and reactive power control.



\bibliographystyle{IEEEtran.bst}
\bibliography{reference}

\end{document}